\documentclass{PoS}

\title{Recent Results on Two-photon Physics at BABAR}

\ShortTitle{Recent Results on Two-photon Physics at BABAR}

\author{\speaker{V.~P.~Druzhinin}
\thanks{On behalf of the BABAR Collaboration}\\
        Budker Institute of Nuclear Physics, Novosibirsk, Russia\\
        E-mail: \email{druzhinin@inp.nsk.su}}

\abstract{Recent BABAR results on two-photon processes are presented.
A  high statistics study of the two-photon production of the charmonium states 
$\eta_c$ and $\eta_c(2S)$ is performed. The mass and width of $\eta_c$ and $\eta_c(2S)$
are measured; the ratio of the decay probabilities to $K_S K^+\pi^-$ and 
$K^+K^-\pi^+\pi^-\pi^0$ are determined. The latter mode is studied for the first time.
The reactions $e^+e^- \to e^+e^- \gamma^\ast \gamma^\ast \to e^+ e^- +$ pseudoscalar meson are
studied in the single-tag mode for $\pi^0$, $\eta$, $\eta^\prime$, and $\eta_c$.
From the measured differential cross sections the $Q^2$ dependencies of the photon-meson
transition form factors are extracted. From these measurements we conclude
that the pion distribution amplitude strongly differs from the distribution amplitudes
of $\eta$ and  $\eta^\prime$ mesons.}

\FullConference{35th International Conference of High Energy Physics - ICHEP2010,\\
		July 22-28, 2010\\
		Paris, France}

\begin{document}

\section{Introduction}
Two-photon production of a resonance $R$ is studied at $e^+e^-$ colliders
in the process $e^+e^-\to e^+e^- R$.
The electrons in this process are scattered 
predominantly at small angles. For the  pseudoscalar meson production, 
the effect of strong interactions is described by only one form factor 
$F(q_1^2, q_2^2)$ depending on the squared momentum transfers to the electrons.

Two-photon processes are usually studied in so called no-tag mode
with both final electrons undetected. In this case the virtual photons emitted by  
electrons are practically real, the momentum transfers squared are close to zero.
In no-tag mode the meson-photon transition form factor at zero $q^2$'s and the
meson two-photon width are measured.

In the single tag-mode one of the final electrons is detected.
The corresponding virtual photon is highly off-shell. From the measurement of
the cross section richer information is extracted: the dependence of
the meson form factor on $Q^2=-q_1^2$.

In this report we present results of no-tag and
single-tag measurements performed with the BABAR detector
at the PEP-II $e^+e^-$ collider. The results are based on data with integrated
luminosity of about 500 fb$^{-1}$ collected at the center-of-mass energy of
10.6 GeV.

No-tag two-photon events are selected by the requirement that the transverse 
momentum of detected hadron system is low. 
The single-tag events are selected with the detected and identified electron
and with the fully reconstructed pseudoscalar meson,  $\pi^0$, $\eta$, $\eta^\prime$, 
or $\eta_c$. It is required that the transverse momentum of the electron-plus-meson 
system be low and the missing mass in an event be close to zero.

\section{Measurement of $\eta_c$ and $\eta_c(2S)$ parameters in the no-tag mode}
\begin{figure}
\includegraphics[width=.49\textwidth]{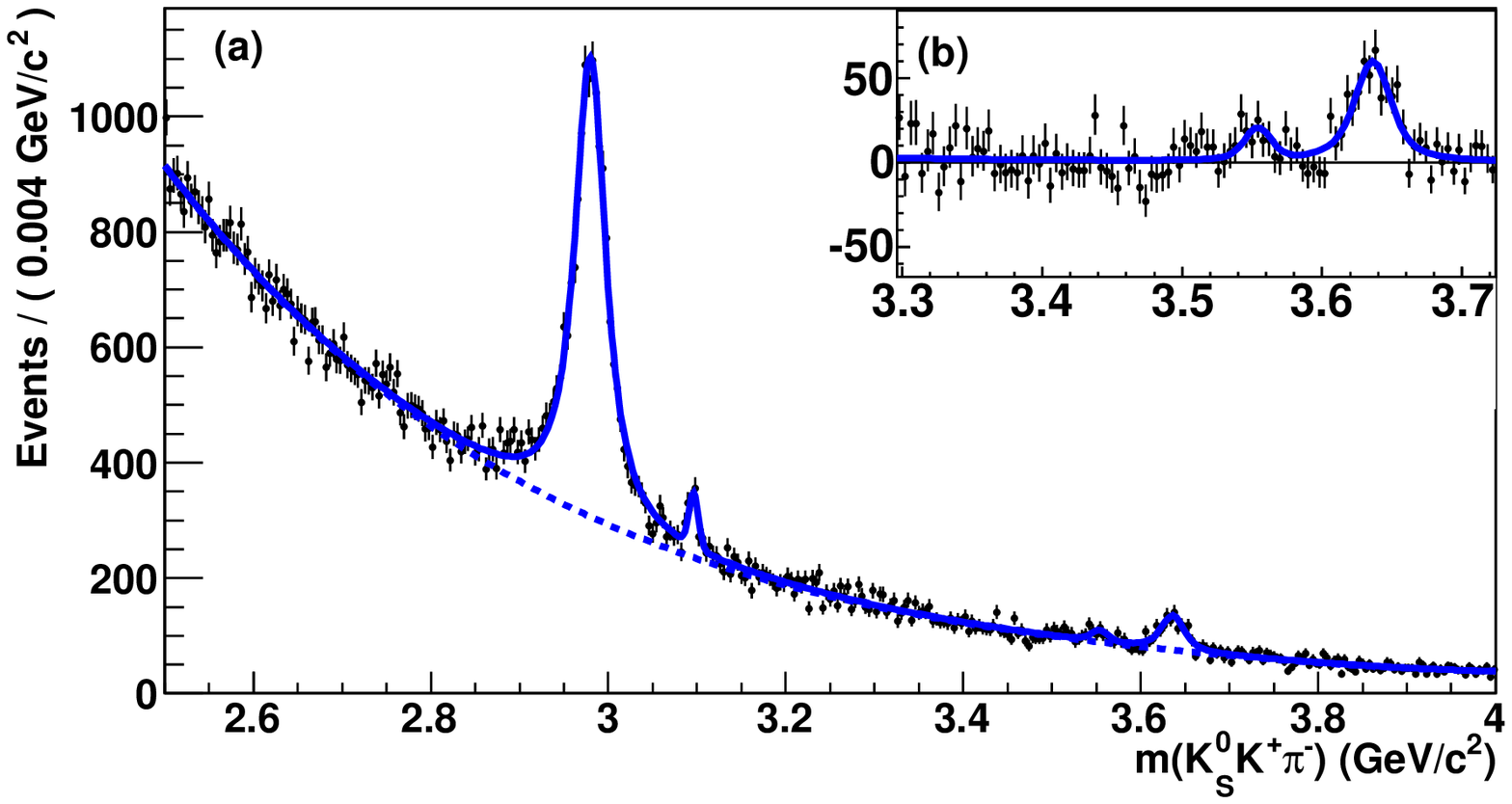}
\hfill
\includegraphics[width=.49\textwidth]{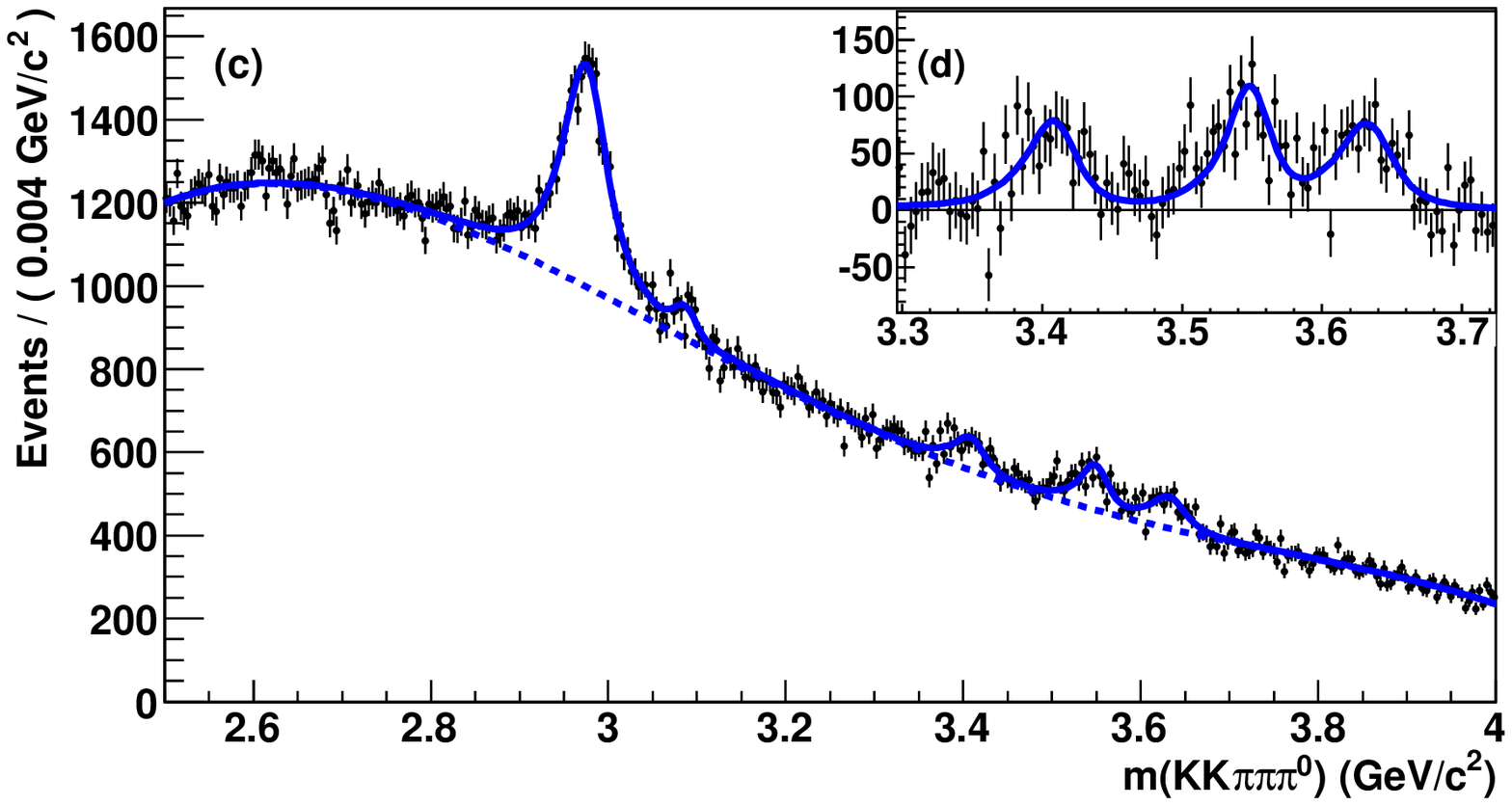}
\caption{The $K_S K^\pm\pi^\mp$ (a) and $K^+K^-\pi^+\pi^-\pi^0$ (c) mass spectra.
The solid line is the fit result. The dashed line represents non-resonant background.
The plots (b) and (d) show background subtracted spectra for the mass range
3.3--3.7 GeV/$c^2$.
\label{fig1}}
\end{figure}

The $K_S K^\pm\pi^\mp$ mass spectrum for no-tag events is shown in Fig.~\ref{fig1}(a).
The $\eta_c$, $J/\psi$, and $\eta_c(2S)$ peaks are clearly seen over a non-resonant smooth
background. The $J/\psi$'s are produced in the initial state radiation process 
$e^+e^-\to J/\psi\gamma$.  
An evidence for the $\chi_{c2}\to K_S K^\pm\pi^\mp$ decay  is also seen
in Fig.~\ref{fig1}b. 
From the fit to the mass spectrum the following $\eta_c$ parameters
are determined~\cite{etacff}:
\begin{eqnarray}
&&m = 2982.2\pm0.4\pm1.5\mbox{ MeV/$c^2$},\,\, \Gamma = 31.7\pm1.2\pm0.8 \mbox{ MeV},\\
&&\Gamma(\eta_c \to \gamma\gamma)B(\eta_c \to K\bar{K}\pi) =
0.379\pm0.009\pm0.031\mbox{ keV}.
\end{eqnarray}
These are the most precise measurements of the $\eta_c$ mass and width to date.
The obtained value of
$\Gamma(\eta_c \to \gamma\gamma)B(\eta_c \to K\bar{K}\pi)$
agrees with the CLEO measurement $0.407\pm0.022\pm0.028$
keV~\cite{CLEO1}.

From the fit to the $K_S K^\pm\pi^\mp$ mass spectrum in the vicinity of the $\eta_c(2S)$
resonance the following values of the $\eta_c(2S)$ mass and width are obtained:
\begin{equation}
m = 3638.3\pm1.5\pm0.6\mbox{ MeV/$c^2$},\,\,\Gamma = 14.2\pm4.4\pm2.5 \mbox{ MeV}.
\end{equation}
These results are preliminary. They  are in reasonable agreement with the previous 
BABAR measurements~\cite{bb_etac}:
$m=3630.8\pm3.4\pm1.0$ MeV/$c^2$ and $\Gamma=17.0\pm8.3\pm2.5$
MeV, obtained using 88 fb$^{-1}$ data. The current PDG values for these
parameters are $m=3637\pm4$ MeV/$c^2$ and $\Gamma=14\pm7$ MeV~\cite{pdg}.
The measured value of the $\eta_c(2S)$ width is also in good agreement with an 
estimation based on a quark model: 
$\Gamma(\eta_c(2S)\to gg)\approx
\Gamma(\eta_c(1S)\to gg)\Gamma(\psi(2S)\to ee)/\Gamma(\psi(1S)\to ee)=
12.1\pm1.0$ MeV.

The mass spectrum for $K^+K^-\pi^+\pi^-\pi^0$ two-photon events is shown in 
Fig.~\ref{fig1}(c). The signals from $\eta_c$, $\chi_{c0}$, $\chi_{c2}$, and 
$\eta_c(2S)$ are seen. This is a first observation of the $K^+K^-\pi^+\pi^-\pi^0$ decay
for these resonances. The $\eta_c(2S)$ meson was previously
observed in only $K_S K\pi$ decay mode. We have determined the ratios of 
the branching fractions into the two decay modes for $\eta_c$ and $\eta_c(2S)$:
\begin{eqnarray}
&&B(\eta_c\to K^+K^-\pi^+\pi^-\pi^0)/B(\eta_c\to K_S K^\pm\pi^\mp)=1.42\pm0.06\pm0.26,\\
&&B(\eta_c(2S)\to K^+K^-\pi^+\pi^-\pi^0)/B(\eta_c(2S)\to K_S K^\pm\pi^\mp)=2.1\pm0.4\pm0.5.
\end{eqnarray}
These results are preliminary.

\section{Measurement of meson-photon transition form factors}
In perturbative QCD, at large $Q^2$, the meson-photon transition form factor can 
be represented as a convolution of a calculable amplitude for $\gamma\gamma^\ast\to
q\bar{q}$ with a nonperturbative meson distribution amplitude (DA)~\cite{LB}. The latter
describes the transition of the meson into two quarks.

Due to the relatively large $c$-quark mass, the $\eta_c$
form factor is rather insensitive to the shape of the $\eta_c$ distribution
amplitude. Its $Q^2$ dependence is expected to be described
by a monopole function with a pole parameter $\Lambda\sim 10$
GeV$^2$~\cite{th1}.  This value is close to the VDM prediction: $\Lambda=m^2_{J/\psi}=9.6$ GeV$^2$. 

The BABAR data on the $Q^2$ dependence of the normalized
$\gamma\gamma^\ast\to\eta_c$ transition form factor~\cite{etacff} is
fitted well by a monopole function. The found pole parameter 
$\Lambda=8.5\pm0.6\pm0.7\mbox{ GeV}^2$ is in agreement with both VDM and QCD predictions,
and with the result of the lattice QCD calculation: 
$\Lambda=8.4\pm0.4\mbox{ GeV}^2$~\cite{lqcd}.
\begin{figure}
\includegraphics[width=.33\textwidth]{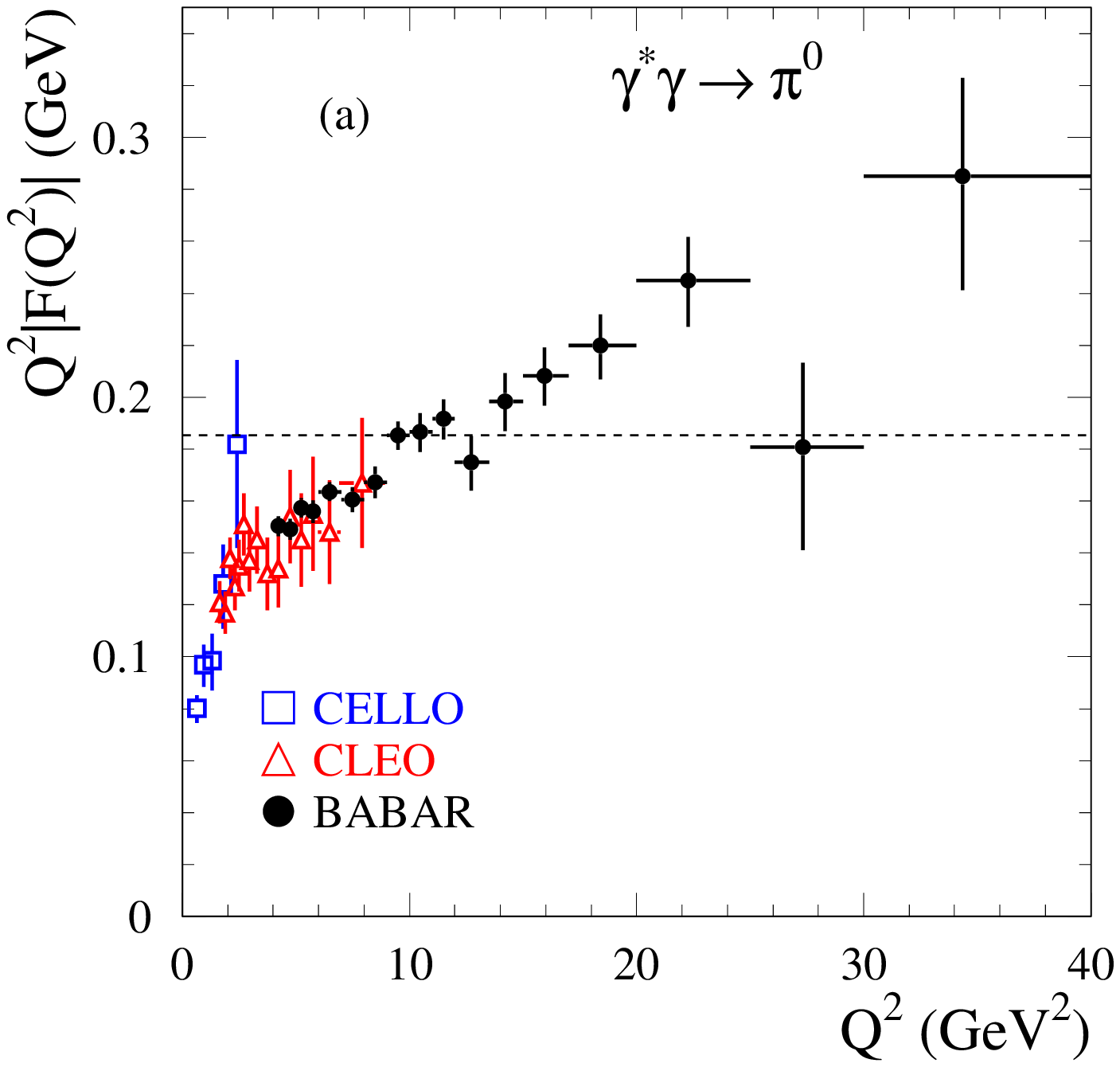}
\includegraphics[width=.33\textwidth]{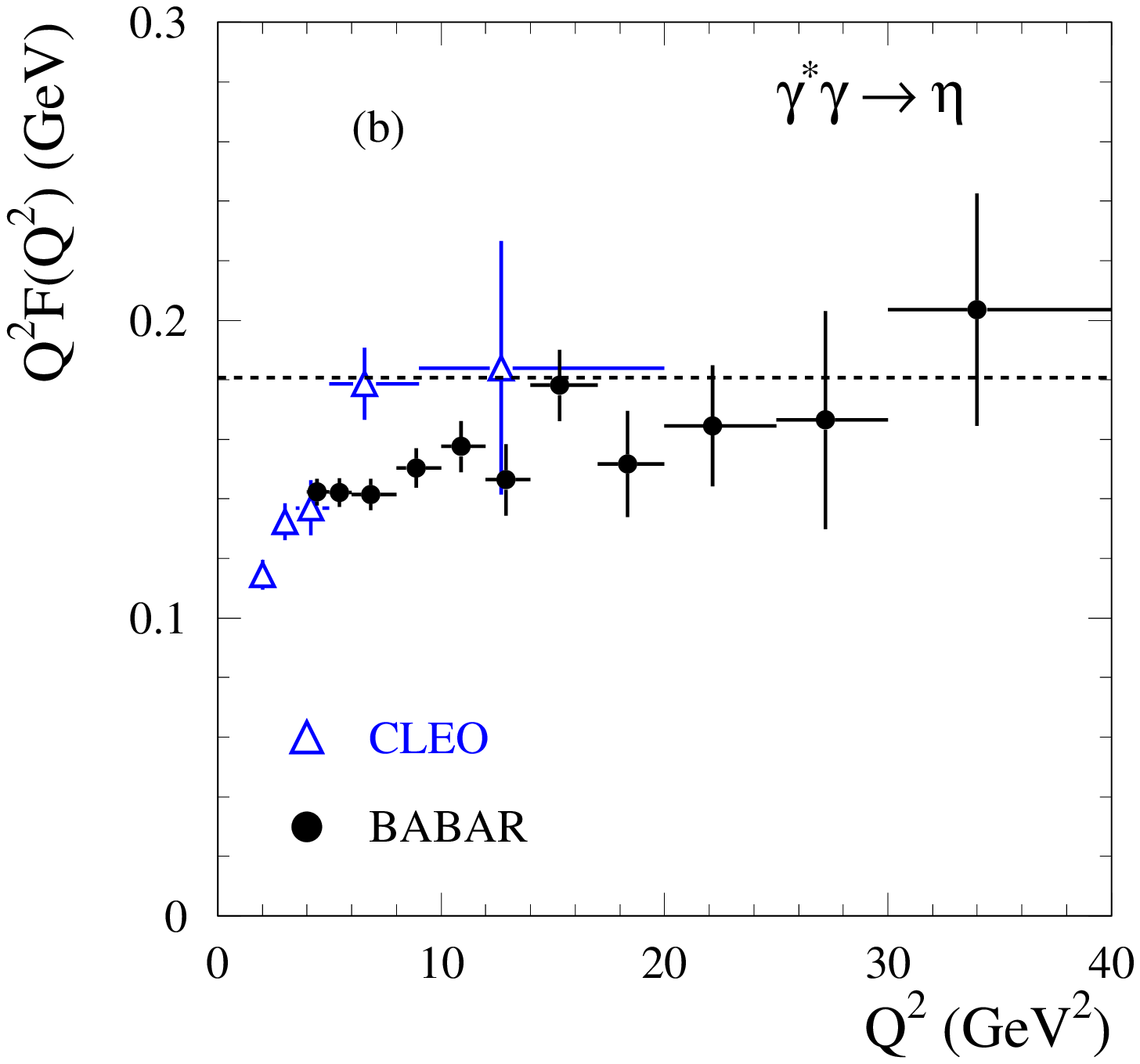}
\includegraphics[width=.33\textwidth]{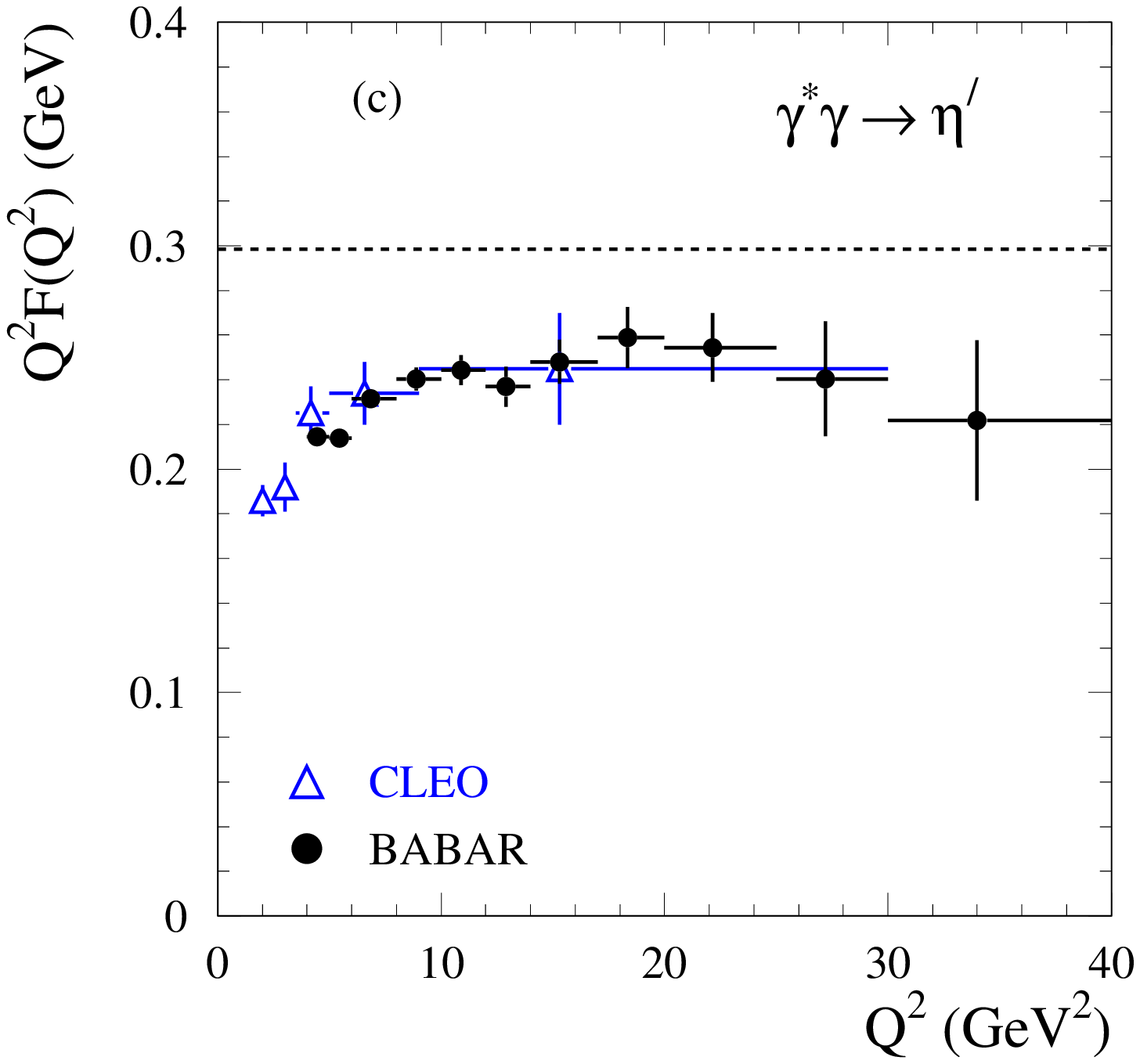}
\caption{The scaled photon-meson transition form factors for $\pi^0$ (a), $\eta$ (b), 
and $\eta^\prime$ (c) mesons. The dashed lines indicate  the asymptotic limits for the scaled
form factors.
\label{fig2}}
\end{figure}

For light pseudoscalars, the form factor depends strongly on the shape
of the meson DA. Experimental data can be used to test different DA models.
The BABAR results~\cite{pi0ff} on the scaled (multiplied by $Q^2$) 
$\gamma\gamma^\ast\to\pi^0$ transition form factor is shown in Fig.~\ref{fig2}(a)
together with CLEO and CELLO data~\cite{CELLO,CLEO}. The horizontal dashed line 
indicates the asymptotic limit for the scaled form factor
($Q^2F(Q^2)=\sqrt{2}f_\pi\approx 0.185$ GeV) predicted by pQCD~\cite{LB}.
The measured form factor exceeds the asymptotic limit at $Q^2 > 10$ GeV$^2$.
This means that the pion DA is significantly wider than the asymptotic DA.
The models with wide or very wide, flat DA's were proposed (see, for example,
Refs.~\cite{after1,after2,after3,after4,after5,after6,after7}) to
describe the $Q^2$ dependence of the pion form factor observed by BABAR.

The BABAR preliminary results on the scaled $\gamma\gamma^\ast\to \eta$ and 
$\eta^\prime$ transition form factors measured in the $e^+e^-\to e^+e^-\eta^{(\prime)}$
reactions are shown in Figs.~\ref{fig2}(b) and (c) in comparison with previous CLEO 
measurements~\cite{CLEO}. We significantly improve the precision and extend the $Q^2$
region for form factor measurements. For $\eta^\prime$ our results and CLEO
measurements are in good agreement. For $\eta$ the agreement is worse. The CLEO
point at 7 GeV$^2$ lies higher than our data by about 3 sigmas.

The $e^+e^- \to \eta^{(\prime)}\gamma$ reactions also can be used to determine
the transition form factors, but in the time-like region $q^2 = s > 0$. The
time- and space-like form factors are expected to be close to each other at high $Q^2$.
The form factors at $Q^2=14.2$ GeV$^2$ 
\begin{equation}
Q^2 F_\eta(Q^2)=0.187\pm0.030\mbox{ GeV},\,\,
Q^2 F_{\eta^\prime}(Q^2)=0.222\pm0.035\mbox{ GeV}
\end{equation}
are obtained from the values of the 
$e^+e^- \to \eta^{(\prime)}\gamma$ cross sections measured by CLEO~\cite{etaff_4} near
the maximum of the $\psi(3770)$ resonance.  The assumption is used that the
contributions of the $\psi(3770) \to \eta^{(\prime)}\gamma$ decays to the
$e^+e^- \to \eta^{(\prime)}\gamma$ cross sections are negligible.
The time-like form factors at $Q^2=14.2$ GeV$^2$ are close to the corresponding space-like
values. The BABAR measurements of the $e^+e^- \to \eta^{(\prime)}\gamma$ cross
sections~\cite{etaff_5}  near the maximum of the $\Upsilon$(4S) resonance 
allows us to extend the $Q^2$ region for the $\eta$ and $\eta^\prime$ form factor
measurements up to 112 GeV$^2$. The time-like form-factor values at 112 GeV$^2$
are as follows:
\begin{equation}
Q^2 F_\eta(Q^2)=0.229\pm0.031\mbox{ GeV},\,\,
Q^2 F_{\eta^\prime}(Q^2)=0.251\pm0.021\mbox{ GeV}.
\end{equation}

The dashed lines in Figs.~\ref{fig2}(b) and (c) indicate the asymptotic limits for 
the scaled $\eta$ and $\eta^\prime$
form factors calculated in Ref.~\cite{Kroll_asy}. It is seen that $Q^2$ dependencies
of the form factors for $\eta$ and $\eta^\prime$ differ from that for $\pi^0$.
We conclude that BABAR results on the meson-photon transition form factors for
light pseudoscalars indicate that the pion DA is significantly wider than the
DA's of $\eta$ and $\eta^\prime$ mesons.

\end{document}